\documentstyle[preprint,aps,prc]{revtex}
\begin{document}

\draft
\title{Nuclear Matter Properties of the Modified Quark Meson Coupling Model}
\author{H.~M\"uller and B.K.~Jennings}
\address{TRIUMF, 4004 Wesbrook Mall, Vancouver, B.C. Canada V6T 2A3}
\vskip1in
\date{\today}
\maketitle
\begin{abstract}
We explore in more detail the modified quark meson coupling (MQMC)
model in nuclear matter. Based on previous studies two different
functional forms for the density dependence of the bag constant are discussed.
For uniform matter distributions the MQMC model can be cast
in a form identical to QHD by a redefinition of the sigma meson
field. It is then clear that modifications similar to those introduced
in QHD will permit the reproduction of all nuclear matter properties
including the compressibility.
After calibrating the model parameters at equilibrium nuclear
matter density, the model and parameter dependence of the resulting
equation of state is examined. Nucleon properties and scaling
relations between the bag constant and the effective nucleon mass are
discussed.
\end{abstract}
\vspace{20pt}
%
%
\section{Introduction}
The description of the nuclear many-body problem in terms of strongly 
interacting quarks and gluons is one of the major challenges in nuclear 
physics.
At present, however, rigorous studies of QCD are restricted to matter systems
at high temperature and zero baryon density. Because of the nonperturbative 
features, it appears very difficult to derive from this theory predictions 
for processes at energy scales relevant for low- and medium-energy nuclear 
phenomenology.

On the other hand,
it is well known that despite these difficulties nuclear phenomenology can be 
efficiently described using hadronic degrees of freedom.
While this framework has been very successful in describing the features of
nuclear matter and the binding energy systematics of finite nuclei,
experiments, such as deep inelastic scattering off nucleons,
provide evidence that the standard hadronic picture has to be corrected. 
For example, the prominent EMC effect which reveals medium 
modifications of the internal structure of the nucleon \cite{ARNEODO94}.

Moreover, hadronic models are often extrapolated into regimes of high 
density and temperature to extract the nuclear equation of state,
which is the basic ingredient in many astrophysical 
applications and in microscopic models of energetic nucleus--nucleus 
collisions. One can expect that under these extreme conditions quark degrees 
of freedom become important.

To address these issues it is necessary to build theories which incorporate
quark-gluon degrees of freedom and which help to bridge the gap 
between nuclear phenomenology and the underlying physics of strong 
interactions. An important criteria for these new models is that they 
reproduce results based on the established hadronic framework.

Guided by the symmetry breaking pattern of QCD,
much effort has recently been devoted to the study of effective Lagrangians
for low-energy strong interactions. 
A typical example is the Nambu and Jona-Lasinio (NJL) model \cite{NJL61}.
These models are mainly concerned with
the spontaneous breaking and restoration of chiral symmetry but it is not
clear if basic features of nuclear phenomenology, such as saturation
of nuclear matter, can be described properly.
On the other hand, almost a decade ago, Guichon \cite{GUICHON88} proposed 
a quark-meson coupling (QMC) model in which
nucleons arise as MIT bags interacting through meson mean fields.
This model was refined later by including center-of-mass corrections 
\cite{FLECK90} and applied to nuclear matter 
\cite{SAITO92,SAITO94,SAITO,SAITO95,THOMAS,SONG95,JIN96a,JIN96b} 
and also, more recently, 
to finite nuclei \cite{GUICHON96,BLUNDEN96}

Although it provides a simple and attractive framework to describe nuclear 
systems in terms of quark degrees of freedom, the QMC model has a serious
shortcoming. It predicts much smaller scalar and vector potentials than 
obtained in successful hadronic models \cite{JIN96a,JIN96b}.
As a consequence the nucleon mass is much too high 
\cite{SAITO92,SAITO94,JIN96a,JIN96b} and the 
spin-orbit force is too weak to explain spin-orbit splittings and spin
observables in finite nuclei.

A well established framework for relativistic hadronic models is provided
by quantum hadrodynamics (QHD) \cite{SEROT97}.
Numerous calculations have established that relativistic mean-field models 
based on QHD provide a realistic description of the 
bulk properties of finite nuclei and nuclear matter \cite{SEROT97}.
One of the key observations in their success is that nucleon propagation
in the nuclear medium is described by a Dirac equation featuring large scalar
and vector potentials. They emerge from the interaction of a nucleon
with all other nucleons in the Fermi sea via the exchange of isoscalar scalar
and vector mesons.

Recently, it was pointed out that the small vector and scalar potentials
in the QMC model are due to the assumption that the bag constant does not
change in the nuclear environment \cite{JIN96a,JIN96b}. By introducing a 
density dependent
bag constant it was demonstrated that large scalar and vector potentials
can be produced.
A necessary condition is
that the value of the bag constant in the nuclear environment
significantly drops below its free-space value. As a consequence relativistic
nuclear phenomenology can be recovered from a modified quark-meson coupling
(MQMC) model \cite{JIN96a}. 

The central issue of the MQMC model is the density dependence of the bag 
constant.
A priori this is not known and the idea of the MQMC model is to parametrize
the bag constant and to determine the parameters by calibrating to 
observed nuclear properties.
Two different model types have been proposed \cite{JIN96a,JIN96b}:
a direct coupling model in which the bag constant is a function of the
scalar field and a scaling model in which the bag constant is related
to the effective nucleon mass.
The density dependence is then generated self-consistently in terms of 
these in-medium quantities.
However, the proposed models are not flexible enough to predict
nuclear matter properties on a satisfactory level.
The trend in the MQMC model is to predict reasonable
values for the effective nucleon mass but compressibilities which are too
high and vice versa \cite{JIN96a,JIN96b}.

We adopt the approach of Refs.~\cite{JIN96a,JIN96b}
with the goal of generalizing the proposed models.
We assume two different functional forms for the bag constant.
In one case it depends on the scalar field only;
in our second model we study the bag constant as a function of the effective 
nucleon mass.
To provide sufficient flexibility we model the functional form by using
polynomial and Pad\a'e parametrizations. The unknown coefficients serve
as our model parameters.
We investigate how well the models can be calibrated by fitting the 
parameters to properties of nuclear matter, which we take
to be: the equilibrium density and binding energy 
($\rho^{\scriptscriptstyle 0}_{\scriptscriptstyle N}$,
$-e_{\scriptscriptstyle 0}$),
the nucleon effective (or Dirac) mass at equilibrium 
($M_{\scriptscriptstyle N,0}^*$)
and the compression modulus ($K_{\scriptscriptstyle 0}$).
In general the models contain more parameters than there are normalization
conditions. 
Thus families of models can be generated
which describe exactly the same nuclear matter properties at equilibrium.
This allows us to study different physical situations and to
search for model dependence in the predictions.

Because the QMC model was proposed to describe ``new'' physics beyond
the standard hadronic picture in nuclear matter and finite nuclei, we
address the crucial question if the model is consistent with established
results.
As a first step we investigate the connection between the MQMC model
and QHD.
We demonstrate that in nuclear matter the MQMC energy functional is formally
equivalent to the expression obtained in QHD with a general nonlinear
scalar potential. 
The quark substructure is entirely contained in the
scalar potential and, in principle, this provides a tool to generate
hadronic potentials based on a quark model.
The explicit form of the hadronic potential depends on the model which
is employed for the bag constant. Thus determining the parameters on the
quark level is equivalent to calibrating the potential on the hadronic
level which is a well established procedure in nuclear matter calculations
\cite{FST96,BODMER91}.

We apply our model to symmetric nuclear matter and compare the predictions
with QHD mean-field calculations. We employ a version of QHD which
contains quartic and cubic scalar self-interactions \cite{BOGUTA77} and 
which is calibrated
to reproduce the same nuclear matter properties as the MQMC model. 
We show that different parametrizations and models for the bag constant lead
to equivalent nuclear properties at low and moderate densities. In this
region the MQMC model predictions are in excellent agreement with QHD. 
In contrast, the original version of the QMC model leads to
substantially different results.
Hence we are lead to the satisfactory conclusion that our generalized
MQMC model predicts nuclear matter properties of the same quality as other
established mean-field models.
The key to this success is the correlation between the bag constant
and nuclear matter properties. Confirming the results of 
Refs.~\cite{JIN96a,JIN96b}, we find that the bag constant has to be 
significantly smaller
than its free-space value to reproduce the desired effective nucleon mass at
equilibrium.
More importantly, the explicit value of the bag constant at equilibrium 
is model independent, {\em i.e.} independent of the details of the 
parametrization.

The exact density dependence of the bag constant is essentially unknown.
However, based on general theoretical arguments scaling relations among
in-medium quantities have been proposed in the recent literature 
\cite{BROWN91,ADAMI93}.
Since our approach establishes a direct connection between nuclear 
phenomenology and the quark substructure we investigate to what extent
the proposed scaling relation between the bag constant and the effective 
nucleon mass is realized; at this point we encounter a clear model
dependence. 
Although all our models are calibrated to produce identical
nuclear matter properties we can build models which exactly follow the
proposed scaling relation but also models which differ considerably.
A similar model dependence can be observed for the bag radius.
In accordance with previous studies \cite{JIN96a,JIN96b} 
we find a sizable increase of the bag radius as a consequence of 
the decreasing bag constant.
However, the quantitative predictions exhibit a strong model and parameter
dependence.

The outline of this paper is as follows:
In Sec.~II, we give a short description of the QMC model and 
and summarize the relations which determine the nuclear matter properties.
Section III is devoted to discuss the connection between the QMC model and
QHD. We also discuss the calibration procedure.
In Sec.~IV, we apply our model to symmetric nuclear matter.
We compare our results with QHD and with the original version of the QMC model.
Section V contains a short summary and our conclusions.
%
%
\section{The Quark-Meson Coupling Model}
In this section, we briefly summarize the relations which determine the 
nuclear equation of state in the quark-meson coupling model. 
For further details we refer the reader to Refs.~\cite{SAITO94,JIN96a,JIN96b}.

In the QMC model the nucleon in the nuclear medium is described as a
static, spherical MIT bag in which quarks couple to meson mean fields.
In symmetric matter they are taken to be neutral scalar $(\sigma) $ and vector 
fields $(V^{\mu})$.

The energy of a bag consisting of three quarks in the ground state can 
be expressed as
\begin{eqnarray}
E_{bag} = 3 {\Omega_q\over R}-{Z\over R}+{4\over 3}\pi R^3 B  \ . 
 \label{eq:ebag}
\end{eqnarray}
where the parameter $Z$ accounts for the zero point motion and $B$ is the
bag constant.
The coupling of the quarks to the scalar field is inherent in the quantities
$\Omega_q$ and $x$ which are given by
\begin{eqnarray}
\Omega_q &=& \sqrt{x^2+(R m_q^*)^2} \nonumber\\
j_0(x)&=&\left({\Omega_q-R m_q^*\over \Omega_q+R m_q^*}\right)^{1/2}j_1(x) \ ,
\label{eq:bessel}
\end{eqnarray}
and where $m_q^* = m_q^0-g_{\sigma}^q\sigma$ denotes the effective 
quark mass and 
$m_q^0$ is the current quark mass.  
For simplicity we work in the chiral limit, {\em i.e.} $m_q^0=0$.

After correcting for the spurious center of mass motion, the effective mass
of a nucleon bag is given by\cite{FLECK90}
\begin{eqnarray}
M_N^* = \sqrt{ E_{bag}^2-3 x^2/R^2 } \ . \label{eq:mstar}
\end{eqnarray}
For a fixed meson mean-field configuration
the bag radius $R$ is determined by the equilibrium
condition for the nucleon bag in the medium
\begin{eqnarray}
{\partial M_N^* \over \partial R} = 0 \ . \label{eq:equilibrium}
\end{eqnarray}
In free space $M_N$ can be fixed
at its experimental value 939 MeV and the condition 
Eq.~(\ref{eq:equilibrium}) to determine the parameters $B=B_0$ and $Z=Z_0$.
For our choice, $R_0=0.6$ fm, the result for $B_0^{1/4}$ and $Z_0$ are
188.1 MeV and 2.03, respectively. 

The total energy density of nuclear matter can be written as 
\cite{FLECK90,SAITO94}
\begin{eqnarray}
{\cal E}_{QMC} &=& 
  {\gamma\over 2\pi^2}\int_{0}^{k_{\rm F}}dk\, k^2(k^2+M_N^*{}^2)^{1/2}
   + g_{\rm v} V_0\rho_N
   - {1\over 2} m_{\rm v}^2 V_0^2
   + {1\over 2} m^2_{\rm s}\sigma^2
   \ .        \label{eq:energybag}
\end{eqnarray}
In symmetric matter ($\gamma=4$) the Fermi momentum of the nucleons
is related to the conserved baryon density by
\begin{eqnarray}
\rho_N={\gamma\over 6\pi^2}k_{\rm F}^3 \ .
         \label{eq:baryondensity}
\end{eqnarray}
The meson mean fields are determined by the general thermodynamic condition
that they should make the energy per nucleon stationary. 
For the time-like component of the vector field this leads to the relation
\begin{eqnarray}
g_{\rm v} V_0
= {g_{\rm v}^2 \over m_{\rm v}^2} \rho_N \ ,
\label{eq:vector}
\end{eqnarray}
whereas the scalar field is determined by 
the self-consistency equation
\begin{eqnarray}
\sigma= {C_q(\sigma)\over m_{\rm s}^2}
{\gamma M_N^*\over 2\pi^2}
\int_{0}^{k_{\rm F}}dk\, {k^2\over(k^2+M_N^*{}^2)^{1/2}} \ .
\label{eq:selfcbag0}
\end{eqnarray}
The details of the quark substructure are entirely contained in the
effective coupling $C_q(\sigma)$ which is related to the effective nucleon mass
by
\begin{eqnarray}
C_q(\sigma)=
-{\partial M_N^* \over \partial \sigma} \ ,
\label{eq:effcoup}
\end{eqnarray}
and which depends on the explicit form of the bag constant. Details can be 
found in Refs.~\cite{JIN96a,JIN96b}.
%
%
\section{GENERATING NONLINEAR MEAN-FIELD MODELS FROM MODIFIED
         QUARK-MESON COUPLING MODELS}
In the original version of the QMC model\cite{GUICHON88,FLECK90,SAITO94}
the bag parameters $B$ and $Z$
were held fixed at their free space values $B=B_0, Z=Z_0$.
The bag constant $B$ is a nonuniversal quantity associated with the QCD trace 
anomaly.
In the nuclear environment it is expected to decrease with increasing density
as argued in Ref.~\cite{ADAMI93}.
At present, however, no reliable information on the medium dependence of 
$B$ is available on the level of QCD calculations.
On the other hand, much effort has recently been devoted to the study of 
effective models which approximate low energy QCD. The guiding principle for
constructing such effective models are the symmetries and the symmetry breaking
patterns of QCD, in particular chiral symmetry breaking. Typical 
representatives are Nambu and Jona-Lasinio (NJL) models \cite{NJL61} 
and related 
chiral meson lagrangians \cite{EBERT86}. In this framework attempts have 
been made to model the QCD trace anomaly by introducing a scalar glueball 
field \cite{SCHECHTER80}.
The concept of a bag constant arises naturally in these models and it is
a common feature to predict a decreasing value of $B$ when the density (or
temperature) of the nuclear environment is increased \cite{ASAKAWA89}.

To account for this physics in the QMC approach two different models 
for the bag constant have been proposed \cite{JIN96a,JIN96b}.
A direct coupling between the bag constant and the scalar mean field
\begin{eqnarray}
{B\over B_0}=
    \left[1-g^B_{\sigma} {4\over \delta} {\sigma\over M_N}\right]^{\delta} \ ,
\label{eq:dcmodel0}
\end{eqnarray}
and a scaling model which relates the bag constant in the medium directly to
the effective nucleon mass
\begin{eqnarray}
{B\over B_0}=
    \left[{M_N^* \over M_N}\right]^{\kappa} \ .
\label{eq:smodel0}
\end{eqnarray}
Thus, rather than focus on the calculation of the bag constant in the medium
the idea of the modified QMC model is to parametrize the density 
dependence in terms of in-medium quantities. A similar approach is used to
construct
nonlinear mean-field models in quantum hadrodynamics (QHD) where the unknown
density dependence of the nuclear energy functional is parametrized by nonliner
meson-meson interactions \cite{FST96,MUELLER96}.
This task is certainly more difficult on the quark level. The direct coupling 
model is inspired by NJL type nontopolgical soliton models for the nucleon
\cite{ALKOFER96}, 
where a scalar soliton field is responsible for the binding of three quarks 
to form a nucleon. 
On the other hand, the scaling model is related  to the idea of a
general scaling relation between in-medium quantities as proposed
by Brown and Rho \cite{BROWN91,ADAMI93}.

Generally, the parameter $Z$ may also be modified in the medium 
\cite{AGUIRRE97}. 
However, in contrast to the bag constant, there is less physical intuition how
the medium dependence of $Z$ can be cast in a model. Here we assume that
the density dependence of $Z$ can be disregarded and take $Z=Z_0$. 

The original QMC model with $B=B_0$ has a serious shortcoming. The predicted
mean fields are much smaller than obtained in established relativistic 
mean-field models \cite{FST96,REINHARD86,FPW87,GAMBHIR90,FURNSTAHL93}.
As a consequence, the effective nucleon mass is much too big
\cite{SAITO92,SAITO94}.
Recently, it has been demonstrated that this shortcoming can be significantly 
corrected and that relativistic nuclear-phenomenology can be recovered from 
the models given by Eq.~(\ref{eq:dcmodel0}) and Eq.~(\ref{eq:smodel0}) 
\cite{JIN96a,JIN96b}. 
However, the systematics of the predicted 
nuclear matter properties is still not satisfactory.
The MQMC model produces acceptable values for the
effective nucleon mass but values for the compressibility which are
too high and vice versa \cite{JIN96a,JIN96b}.
In our work we generalize the direct coupling model and the scaling model
demonstrating that the resulting improved MQMC predicts nuclear matter 
properties of the same quality as in other successful relativistic 
mean-field models
\footnote{Experience has shown
that an accurate reproduction of nuclear matter properties leads to realistic
results when the calculations are extended to finite
nuclei \cite{FST96,REINHARD86,FPW87,GAMBHIR90,FURNSTAHL93}}.

Furthermore, we will show that the relation between the QMC model and QHD 
is more general and direct than previously thought \cite{SAITO94,JIN96b}. 
This will allow us a more consistent comparison between these 
different approaches.

We start the analysis with a comparison of the energy density in 
Eq.~(\ref{eq:energybag})
with the corresponding expression in QHD \cite{SEROT97}
\begin{eqnarray}
{\cal E}_{QHD} &=& 
  {\gamma\over 2\pi^2}\int_{0}^{k_{\rm F}}dk\, k^2(k^2+M_N^*{}^2)^{1/2}
   + {g_{\rm v}^2\over 2 m^2_{\rm v}}\rho_N^2
   + U_{\rm s}(\phi)
   \ ,        \label{eq:energyqhd}
\end{eqnarray}
where the effective nucleon mass is given by $M_N^* = M_N- g_s \phi$.
The standard form of the nonlinear scalar potential is 
\begin{eqnarray}
U_{\rm s}(\phi)={1\over 2} m_{\rm s}^2 \phi^2 + 
        {\kappa\over 6} \phi^3 + {\lambda\over 24} \phi^4
   \ .        \label{eq:nlpot0}
\end{eqnarray}
Although the cubic
and quartic terms provide sufficient flexibility
for an accurate calibration of the nuclear equation of state \cite{BOGUTA77},
we will assume a general functional form for the potential.
The scalar field is determined by the self-consistency equation
\begin{eqnarray}
{\partial U_{\rm s}(\phi)\over \partial\phi}=
g_{\rm s}{\gamma M_N^*\over 2\pi^2}
\int_{0}^{k_{\rm F}}dk\, {k^2\over(k^2+M_N^*{}^2)^{1/2}}  \ .
\label{eq:selfcqhd}
\end{eqnarray}
The equivalence of the two sets of equations, Eqs.~(\ref{eq:energybag}) and 
(\ref{eq:selfcbag0}), and Eqs.~(\ref{eq:energyqhd}) and (\ref{eq:selfcqhd}),
can be demonstrated by performing a redefinition of the scalar field in the 
QMC model
\begin{eqnarray}
g_0 \phi (\sigma) \equiv M_N-M_N^*(\sigma) 
                  =  M_N-\sqrt{ E_{bag}^2-3 x^2/R^2 } 
\ . \label{eq:transfb}
\end{eqnarray}
The transformation does not depend explicitly on the density if we assume
\begin{eqnarray}
B=B(\sigma,M_N^*) \ ,
\nonumber
\end{eqnarray}
as suggested in Eqs.~(\ref{eq:dcmodel0}) and (\ref{eq:smodel0}).
The coupling $g_0$ is chosen to normalize the new field according to
\begin{eqnarray}
\phi (\sigma) {\mathop{=}_{\sigma \to 0}} \sigma + O(\sigma^2) \ ,
\nonumber
\end{eqnarray}
and it is given by
\begin{eqnarray}
g_0 = - {\partial M_N^* (\sigma)\over \partial \sigma }\biggr|_{\sigma=0} \ .
\label{eq:geff}
\end{eqnarray}
The transformation is well defined provided that the effective mass
changes monotonically with the scalar field. Inverting the relation
Eq.~(\ref{eq:transfb})
leads to a nonlinear scalar potential in the energy density 
Eq.~(\ref{eq:energybag}) 
\begin{eqnarray}
{1\over 2} m_s^2 \sigma^2(\phi)\equiv
U_{\rm s}(\phi) \ .
\label{eq:nlpotbag}
\end{eqnarray}
At this point the energy density in the QMC model is {\em identical} to 
the QHD expression Eq.~(\ref{eq:energyqhd}) with a general scalar potential. 
For finite nuclei the transformation also introduces a change in the 
kinetic energy of the scalar meson. This will mainly lead to a different 
description of the surface energy, but this effect is expected to be small.

As shown in
Ref.~\cite{JIN96b} a significant simplification arises for models with
$B=B(\sigma)$ if
$g_{\sigma}^q=0$, {\em i.e.} if there is no direct coupling of the quarks 
to the scalar field.
In this case one obtains the scaling relation \cite{JIN96b}
\begin{eqnarray}
{B\over B_0}=\left({M_N^*\over M_N}\right)^4
\label{eq:scaling} \ .
\end{eqnarray}
Thus, if the bag constant is of the specific form
\begin{eqnarray}
{B\over B_0}=\left(1- g_B {\sigma\over M_N}\right)^4 \ ,
\label{eq:simple}
\end{eqnarray}
the transformation Eq.~(\ref{eq:transfb}) is linear
leading to the original Walecka model \cite{WALECKA74} with a
quadratic potential
\begin{eqnarray}
U_{\rm s}(\phi)={1\over 2} m_{\rm s}^2 \phi^2
   \ .   
\end{eqnarray}

The details of the quark substructure are 
entirely contained in the nonlinear potential $U_{\rm s}(\phi)$.
This implies that if the bag constant, more generally the bag parameters, was
known as a function of the scalar field and the effective mass, the steps 
leading to Eq.~(\ref{eq:nlpotbag}) would permit the {\em prediction} 
of potentials for hadronic mean-field models.

On the other hand starting with a known potential, 
Eq.~(\ref{eq:nlpotbag}) defines the inverse field transformation 
to Eq.~(\ref{eq:transfb})
\begin{eqnarray}
\sigma(\phi)\equiv {\sqrt{2 U_{\rm s}(\phi)}\over m_{\rm s}} \ .
\label{eq:invtrans}
\end{eqnarray}
After substituting this result in Eq.~(\ref{eq:energyqhd}), the QHD expression 
for the energy is exactly of the same form as in the MQMC model. 
The self-consistency condition Eq.~(\ref{eq:selfcqhd}) changes to
\begin{eqnarray}
\sigma= {C(\sigma) \over m_{\rm s}^2}
{\gamma M_N^*\over 2\pi^2}
\int_{0}^{k_{\rm F}}dk\, {k^2\over(k^2+M_N^*{}^2)^{1/2}}  \ ,
\label{eq:selfcqhdt}
\end{eqnarray}
where the coupling is given by
\begin{eqnarray}
C(\sigma)=g_s m_{\rm s}
{\sqrt{2U_{\rm s}}\over{\partial U_{\rm s}\over \partial \phi}} \ .
\label{eq:effcoupqhd}
\end{eqnarray}
Thus, the MQMC model is formally equivalent to a nuclear mean field model with
a {\em field dependent} scalar-nucleon coupling.
Field dependent nucleon-meson couplings are often interpreted
as an indication for the compositeness of the nucleon \cite{SAITO94,GUICHON96} 
but from a modern point of view
they appear naturally in hadronic models as a result of field redefinitions.
In our case the field dependent coupling is equivalent to a subset of
nonlinear meson self-interactions
\footnote{
For a discussion of the equivalence between field dependent meson-nucleon 
couplings and nonlinear meson self-interactions 
in the framework of QHD see Ref.~\cite{SEROT97}}.

Using the relation between MQMC and QHD one can determine the bag 
constant as a function of 
$\sigma$.  This approach is useful since
it establishes a direct connection between nuclear phenomenology and the
medium dependence of the bag parameters.

To be more specific let us assume $B=B(\sigma)$ and $g_{\sigma}^q=0$.
The scaling relation Eq.~(\ref{eq:scaling}) together with the expression
for the effective mass in QHD
\begin{eqnarray}
M_N^*=M_N-g_{\rm s} \phi(\sigma)
\end{eqnarray}
then leads to
\begin{eqnarray}
{B\over B_0}=\left(1-{g_{\rm s} \phi(\sigma)\over M_N}\right)^4  \ .
\label{eq:bqhd}
\end{eqnarray}
In general, the transformation which relates the two scalar fields is not known,
but Eq.~(\ref{eq:bqhd}) motivates the {\em ansatz}
\begin{eqnarray}
{B\over B_0}=\left(1-g_B{\sigma\over M_N}F(\sigma)\right)^{\kappa}
\quad \hbox{\rm with} \quad F(0)=1 \ ,
\label{eq:ansatzdc}
\end{eqnarray}
which includes the original form Eq.~(\ref{eq:dcmodel0}).

Similarly, the scaling model can be generalized by 
\begin{eqnarray}
{B\over B_0}=\left({M_N^*\over M_N}\right)^{\kappa} G(M_N^*)
\quad \hbox{\rm with} \quad G(1)=1 \ .
\label{eq:ansatzsc}
\end{eqnarray}
We studied polynomial and Pad\a'e parametrizations for the unknown 
functions $F$ 
and $G$ which all lead to qualitatively similar results at low and moderate
densities. In the following we will present results obtained by using
a simple polynomial
\begin{eqnarray}
F(\sigma)=1+\alpha\sigma+\beta\sigma^2 \ ,
\label{eq:poly}
\end{eqnarray}
for the generalization of the direct coupling model,
and a Pad\a'e form
\begin{eqnarray}
G(M_N^*)={1\over 1-a-b-c
          + a M_N^*+b M_N^*{}^2+c M_N^*{}^3} \ ,
\label{eq:pade}
\end{eqnarray}
for the generalization of the scaling model.
For notational convenience
we will refer to the model in Eq.~(\ref{eq:ansatzdc}) 
as MQMC$_{\rm A}$ and to the model in Eq.~(\ref{eq:ansatzsc}) as MQMC$_{\rm B}$.

Since the parameters $B_0$ and $Z$ are fixed to reproduce the nucleon  mass
in the vacuum our models contain six free parameters. 
The parametrization of the bag constant contains
the parameter $\kappa$ and
the three couplings $(g_B, \alpha, \beta)$ 
and $(a, b, c)$ for MQMC$_{\rm A}$
and for MQMC$_{\rm B}$, respectively;
in addition values for the quark-meson couplings 
$g_{\sigma}^q$ and for the ratio $g_{\rm v}/m_{\rm v}$ are needed
\footnote{Strictly speaking, in nuclear matter
only the ratios $(g_B/g_\sigma^q, \alpha/g_\sigma^q, \beta/g_\sigma^q{}^2,
g_\sigma^q/m_\sigma, g_{\rm v}/m_{\rm v})$ are relevant for
the calibration procedure.
In order to vary the scalar coupling the mass of the scalar meson needs 
to be fixed. We choose $m_\sigma=550$ MeV.}.
Four of the six parameters can be chosen to reproduce the
equilibrium properties of symmetric nuclear matter, which we take as
the equilibrium density and binding energy 
($\rho^{\scriptscriptstyle 0}_{\scriptscriptstyle N}$,
$-e_{\scriptscriptstyle 0}$),
the nucleon effective mass at equilibrium 
($M_{\scriptscriptstyle N,0}^*$) and
the compression modulus ($K_{\scriptscriptstyle 0}$).
The first three of these are tightly constrained \cite{FURNSTAHL93},
whereas the latter is not.
The set of equilibrium properties used here \cite{FST96} are listed
in Table~\ref{tab:one}; these are motivated by successful
descriptions of bulk and single-particle nuclear properties 
\cite{FST96,FURNSTAHL93,FTS95}.
Since there are more free parameters than constraints, we proceed as follows.
We choose values for the coupling $g_{\sigma}^q$ and for $\kappa$
and determine the remaining couplings by requiring that they reproduce
the desired equilibrium properties.
This is achieved by solving a set of transcendental equations
that relate the parameters directly to the nuclear matter
properties; this is a well established procedure in nuclear
matter calculations\cite{FST96,BODMER91}.
The original version of the QMC model with $B=const.$ contains only
two parameters, $g_{\sigma}^q$ and $g_{\rm v}/m_{\rm v}$, which
are usually chosen to reproduce the binding energy and the density at
nuclear matter equilibrium. In our discussion we use the parameters given
in Ref.~\cite{SAITO94}.
For comparison we employ the QHD model with the nonliner potential 
in Eq.(\ref{eq:nlpot0}). The parameters are determined to reproduce the
same equilibrium properties as in the QMC model.

Our primary goal is to study the influence of the in-medium bag constant on
the equation of state (EOS) of nuclear matter.
The freedom of 
varying the parameter $\kappa$ will 
allow us to investigate scaling relations between the bag constant and the 
effective nucleon mass.
Moreover, models with $g_{\sigma}^q = 0$ and $g_{\sigma}^q \neq 0$ describe 
different physical situations. 
For $g_{\sigma}^q = 0$ the scalar meson couples only to the surface of the bag.
Apart from an overall shift in the single-particle energies, due to the 
vector-meson mean 
field, the properties of the quarks inside the bag are not changed by the
nuclear medium.
In contrast, the quarks acquire a density dependent effective mass
$m_q^* = m_q^0-g_{\sigma}^q\sigma$ for nonvanishing couplings $g_{\sigma}^q$.

A decreasing effective nucleon mass in the original QMC model with 
$B=B_0$ requires a positive quark-meson coupling ($g_{\sigma}^q > 0$)
which implies a negative effective quark mass.
A similar effect can be observed in soliton models \cite{ALKOFER96}.
Close to the origin of the soliton the attractive scalar potential leads to a 
negative effective quark mass. But here this is less disturbing since the
effective mass depends on the space coordinates and becomes positive
in the outer region of the soliton.
In the MQMC it is possible to chose $g_{\sigma}^q > 0$ or $g_{\sigma}^q < 0$
and to generate positive or negative effective quark masses.
However, we observe that the sign of the effective quark mass has no
impact 
on the properties of nuclear matter. Models with either sign are
qualitatively equivalent.

A priori we have no specific guidance on the allowed values of $\kappa$ and 
$g_{\sigma}^q$. We observe that not all possible choices permit the 
reproduction of the desired equilibrium properties. For example, MQMC$_{\rm B}$
has no solution for $g_{\sigma}^q \alt 0.8$.
We analyzed MQMC$_{\rm A}$ for
$3\alt \kappa \alt 7$ and $0\leq g_{\sigma}^q \alt 2$,
and MQMC$_{\rm B}$ for
$3\alt \kappa \alt 5$ and $0.8\alt g_{\sigma}^q \alt 1.5$.

%
\section{Nuclear Matter Properties}
The density dependence of the nucleon mass is controlled by the effective
coupling $C_q(\sigma)$ in the self-consistency equation 
Eq.~(\ref{eq:selfcbag0}).
For MQMC$_{\rm A}$ this quantity is indicated in 
Fig.~\ref{fig:effcoup}.
To normalize the curves at the origin we divided $C_q(\sigma)$
by $g_0$ which was defined in Eq.~(\ref{eq:geff}).
The effective coupling for QHD, given by 
Eq.~(\ref{eq:effcoupqhd}), and 
for the original QMC model are also indicated. We emphasize that all 
parametrizations except the original QMC model reproduce the same equilibrium
properties listed in Table~\ref{tab:one}. 
For small values of the scalar field MQMC$_{\rm A}$  and QHD are in 
good agreement predicting nearly constant effective couplings.
In contrast, the curve for the QMC model steadily decreases and 
eventually becomes negative. 

The consequences for the effective
nucleon mass can be studied in Fig.~\ref{fig:mstardc}.
As expected from Fig.~\ref{fig:effcoup}
the masses for 
MQMC$_{\rm A}$ and for QHD are nearly identical up to 1.5
nuclear matter densities. For $B = B_0$ the effective mass decreases very
slowly and, when the point $C_q(\sigma) = 0$ is reached, increases again.
The curves terminate at some maximum density, which corresponds to a
maximum value of the scalar field.
For $g_{\sigma}^q = 0$ this occurs at $B=M_N^*=0$. 
At that point the radius diverges. 
The curves for $g_{\sigma}^q \neq 0$ terminate when $x$, the solution
of Eq.~(\ref{eq:bessel}), approaches zero.

Fig.\ref{fig:mstarsc} shows the effective nucleon mass for MQMC$_{\rm B}$.
Similar to what we observed for MQMC$_{\rm A}$,
the curves are nearly indistinguishable
from the QHD result at low and moderate densities.
At higher densities
MQMC$_{\rm B}$ leads to a slowly decreasing effective mass which approaches
a nonzero asymptotic value.

Fig.\ref{fig:pot} indicates the predicted nonliner potentials in 
Eq.~(\ref{eq:nlpotbag}) as a function of the transformed scalar field
given by Eq.~(\ref{eq:transfb}). The value $g_0\phi/M_N = 0.4$ 
corresponds to the saturation point of nuclear matter.
For $g_0\phi/M_N< 0.6$ the predicted MQMC potentials are almost 
identical to the nonlinear QHD potential in Eq.~(\ref{eq:nlpot0}). In contrast
the potential in the QMC model is much smaller. Here the saturation
point is at $g_0\phi/M_N = 0.11$.

Due to the calibration procedure, 
the very good agreement of the different models and parametrizations at low
and moderate densities is certainly not surprising. 
As clearly visible in Figs.~\ref{fig:effcoup}-\ref{fig:pot},
the predictions of MQMC$_{\rm A}$ and MQMC$_{\rm B}$ vary considerably 
from QHD above the saturation point of nuclear matter.
We also
observe that different model types and different parametrizations within a  
specific model are no longer equivalent at high densities.
This is certainly an indication that an extrapolation 
into the regime of high densities might be problematic. 

In Fig.~\ref{fig:eos} we show the binding energy curves of symmetric matter
for MQMC$_{\rm A}$.
The stiffness of the EOS is controlled by the parameters $\kappa$ and
$g_{\sigma}^q$. 
The curves become softer for smaller
values of $\kappa$ and stiffer if the quark-meson coupling is increased
\cite{JIN96a,JIN96b}.
Also indicated in Fig.~\ref{fig:eos} is the binding energy curve of 
the original QMC model.
Although the compression modulus is only slightly lower ($K_0= 223$MeV)
than in the other models ($K_0= 250$MeV) the EOS in the original QMC
model is substantially softer. To understand this difference, it is useful to
split the binding energy according to
\begin{eqnarray}
e_0={{\cal E}\over\rho_N} -M_N ={{\cal E}^0\over\rho_N} 
 + {U_{\rm v}\over \rho_N} + {U_{\rm s}\over \rho_N} 
   \ ,        \label{eq:sat}
\end{eqnarray}
with
\begin{eqnarray}
{\cal E}^0&=&{\gamma\over 2\pi^2}
 \int_{0}^{k_{\rm F}}dk\, k^2(k^2+M_N^*{}^2)^{1/2} -M_N \rho_N \ , \nonumber\\
 U_{\rm v}&=& {g_{\rm v}^2\over 2m^2_{\rm v}}\rho_N^2 \ , \nonumber \\
\quad \hbox{\rm and} \quad
U_{\rm s}&=&{1\over 2} m^2_{\rm s} \sigma^2 \ .
\nonumber
\end{eqnarray}
In Fig.~\ref{fig:sat} the three different contributions are indicated
as a function of the density. Saturation arises as a competition between
the decreasing effective mass which lowers the kinetic energy and the increase
of the energy due to the increasing density.
As expected, the curves for QHD and MQMC$_{\rm A}$ are almost 
identical. The quantitative difference to the original QMC model is striking.
To achieve saturation at a very high effective mass 
($M_N^*=0.89$) only a
small contribution of the vector part $U_{\rm v}$, {\em i.e.} a small 
coupling $g_{\rm v}^2/ m^2_{\rm v}$, is required. At high 
densities the term $U_{\rm v}$ dominates and a smaller coupling leads 
to a softer EOS. This also explains the tendency of the QMC model to produce
small vector mean fields \cite{JIN96a,JIN96b}.

The in-medium bag constant as a function of the 
effective nucleon mass is indicated in Fig.~\ref{fig:bag}.
We consider MQMC$_{\rm A}$ in part (a) and MQMC$_{\rm B}$ in part (b).
To reproduce the desired effective nucleon mass at the saturation point,
$B$ is required to decrease substantially below its free-space value.
The values of $B/B_0$ are between the two lines $(M_N^*/M_N)^4$
and $(M_N^*/M_N)^3$. 
As mentioned in the last section, in the MQMC$_{\rm A}$ model
a special situation arises for $g_\sigma^q=0$.
In that case the bag constant scales as $(M_N^*/M_N)^4$ for all values 
of $\kappa$ \cite{JIN96b}.

We emphasize that all curves produce identical nuclear matter properties. 
Hence there is no compelling evidence from
nuclear phenomenology that the bag constant should scale with a specific 
power $\kappa$,
{\em e.g.}, like $(M_N^*/M_N)^4$ as proposed in Ref.~\cite{BROWN91}.
The tendency of the MQMC approach is to predict values of $B$ which are slightly
higher. 
This becomes more apparent in Fig.~\ref{fig:bagsat} where the bag 
constant
at the saturation point as a function of $g_q^{\sigma}$ is indicated.
We examine MQMC$_{\rm A}$ for three different values of the
nucleon mass at equilibrium. 
The value $g_q^{\sigma}=0$ correspond to the scaling behavior 
$B/B_0=(M_N^*/M_N)^4$. 
Because the calibration procedure determines
the value of $B$, the curves are {\em model independent}, 
{\em i.e.} they depend only on the coupling but they are independent of the 
functional form of $B=B(\sigma)$. 
The curves in Fig.~\ref{fig:bagsat} are
also independent of the compressibility which merely determines the
derivatives of $B$ with respect to the scalar field in MQMC$_{\rm A}$
and the derivative with respect to the effective nucleon
mass in MQMC$_{\rm B}$. The key quantity here really is the effective 
nucleon mass.

At this point, some caveats concerning the comparison with other approaches
must be added.
As mentioned earlier, the prediction of a reduced bag constant in 
the nuclear environment is also a common feature in effective models
for low energy QCD, {\em e.g.} chiral models. Since none of 
these models can strictly be
derived from QCD it is not clear to what extent results can be compared.
Within the MQMC approach, it was one of our goals to demonstrate that it is 
possible to obtain nuclear matter results which are of the same quality as in
any other successful hadronic mean-field model.
In contrast, chiral models are more concerned with the description of the
underlying physics, {\em i.e.} the breaking and restoration of
chiral symmetry. To the best of our knowledge, it has not been
demonstrated whether it is possible to describe nuclear saturation and
properties of finite nuclei on that level.

The corresponding bag radius is indicated in Fig.~\ref{fig:radius}.
As a consequence of the rapidly decreasing bag constant
the MQMC models predict a picture of a significantly ''swollen'' nucleus
\cite{JIN96a,JIN96b}.
This effect is most drastic in MQMC$_{\rm A}$ for vanishing quark-meson
couplings. As mentioned above, the effective nucleon mass and therefore also
the bag constant vanishes at some finite value of the density. 
Consequently, the radius diverges at this point.
Although all the models produce identical nuclear matter properties
near equilibrium
they yield significantly different radii, even at low densities.
Also shown in Fig.~\ref{fig:radius} is the curve
$R_c=(3/4\pi\rho_N)^{1/3}$ which indicates the
''critical'' radius 
where the individual bags start overlapping, signaling
the breakdown of the simple bag model. In our model this occurs slightly
above the saturation density. 
This behavior is in sharp contrast to the original QMC model which predicts a 
nearly constant radius. 

We emphasize that it is not possible to achieve 
reasonable nuclear matter properties and a small radius at the same time
because the calibration procedure determines the value of the bag constant.
As indicated in Fig.~\ref{fig:bagsat},
the acceptable
range of $M^*_N/M_N=0.6-0.7$ always requires a small value of $B$ which in
turn leads to a large radius. 
This is certainly a dilemma.
On one hand we have demonstrated that the MQMC model can be improved and 
calibrated so that nuclear matter properties are accurately reproduced.
The most important quantity here is the effective nucleon mass which is
tightly constrained by nuclear observables \cite{FURNSTAHL93}.
On the other hand we found that the required bag constant in the medium
leads to radii which are unreasonably high. 
Furthermore the radii are very sensitive to the model features
and parametrizations.
The size of a nucleon in matter is certainly a subtle concept
and one might argue that physical observables do not depend on the
bag radius \cite{BROWN88}. However, it is an important 
phenomenological quantity in many nuclear physics issues where bag models
are employed to describe features which depend on the intrinsic structure
of the nucleon ( see for example \cite{BROWN89}).

Moreover, one motivation for generating the nuclear equation of state 
on basis of a quark
model is the hope that such an approach is well suited to describe systems at
high densities where quark degrees of freedom become important.
However, because of the large radii and the picture of overlapping bags
it is not clear how far the equation of state can be extrapolated into the 
high-density regime. 
%
%
\section{Summary}
In this paper we study properties of nuclear matter based on an improved
quark-meson coupling model.
This model describes nucleons as nonoverlapping MIT bags interacting
through scalar and vector mean fields.
Of central importance is the bag constant which we assume to depend on
the density of the nuclear environment. 
We study two types of models for the bag constant in which the
density dependence is parametrized in terms of in-medium quantities.
In one we assume that the bag constant depends on the scalar field only
and in another the density dependence is related to the effective nucleon mass.

By performing a redefinition of the scalar field we demonstrate that
the resulting energy functional corresponds to a QHD-type hadronic mean-field
model with a general nonlinear scalar potential. In principle, this
connection can be used to generate hadronic potentials from quark models.
Moreover, a direct relation between nuclear phenomenology and the quark
picture arises. We use this connection to motivate our models for the
density dependence of the bag constant.

For the explicit calculations we employ
a polynomial and a Pad\a'e form to model the medium dependence of
the bag constant.
The unknown parameters can then be fit to properties of nuclear matter near
equilibrium that are known to be characteristic of the observed bulk and
single-particle properties of nuclei.

Our basic goal is to study properties of nuclear matter. We investigate
whether the models for the bag constant lead to results which are consistent 
with established hadronic models.
This is relevant in view of the hope to apply quark models to describe
``new'' physics which goes beyond the hadronic picture.
Because of the relation between MQMC and QHD we compare our results
with a QHD model calibrated to produce the same equilibrium properties.
As the basic result we find an excellent agreement between our refined
MQMC models and QHD at low and moderate densities. In particular it is possible
to reproduce the desired saturation properties and large scalar and
vector potentials as demanded by nuclear phenomenology.
The central quantity here is the effective nucleon mass. Its accurate 
reproduction requires a rapidly decreasing bag constant 
below nuclear matter densities which leads to a
satisfactory saturation mechanism. 
In contrast, the original QMC model
predicts a very high effective nucleon mass and saturation is achieved on
much smaller energy scales. 

Moreover, the calibration procedure determines the value of the bag constant
as a function of the effective nucleon mass in a model independent manner. 
However, this does not imply that the density dependence of $B$ is well
constrained.
On the contrary, we find that models which reproduce identical properties
of nuclear matter can be generated from different parametrizations and
models for the bag constant. In particular there is no clear evidence
for a specific scaling behavior of $B$.

We observe a similar model dependence in the predicted bag radius.
As a consequence of the decreasing bag constant the size of the
nucleon significantly increases in all our models. This effect is even
more drastic than in earlier calculations \cite{JIN96a,JIN96b}. 
We find that the nucleon bags
start overlapping at densities slightly above the saturation point
signaling the break down of the model. 
Moreover, different models and parametrizations which are equivalent near
equilibrium produce different high-density equations of states. A similar
picture arises in hadronic mean-field models \cite{MUELLER96}.
Thus, even if quark degrees of freedom are incorporated it is unclear how 
far the nuclear
equation of state can be extrapolated into the high-density regime.

In recent applications the original QMC model was used to estimate the density 
dependence of quark condensates and other hadron masses \cite{SAITO95}. 
In view of our results we expect that the predictions for these quantities
change drastically if the density dependence of the bag constant is taken
into account. It is also important to investigate the uncertainties in these
quantities which arise from the model and parameter dependence 
of the bag constant.

In summary we conclude that an improved QMC model provides a very satisfactory
description of nuclear matter near equilibrium. Phenomenological information
in terms of equilibrium properties of nuclear matter can be used to constrain 
the unknown density dependence of the bag constant. On the hadronic level
different models predict equivalent equation of states at low and moderate
densities. This picture changes if microscopic quantities on the quark level
are considered. Here a clear model and parameter dependence emerges.
The difficulties rely on the detailed density 
dependence of the bag constant. Although these details
have minor impact on the nuclear matter properties, microscopic quantities
like the nucleon radius appear to be very sensitive. It is important to have
these uncertainties under control before one can make reliable statements
about the physics beyond the standard hadronic picture. 
To achieve this goal more detailed empirical information on effective hadron
masses and also additional observables from finite nuclei might provide
useful constrains to reduce model and parameter dependence in the future.

\acknowledgements

We thank T. R. Hemmert for useful comments.
This work was supported by the Natural Science and Engineering Research Council
of Canada.
%
%
\begin{table}[tbhp]
\caption{Equilibrium Properties of Nuclear Matter}
\medskip
\begin{tabular}[b]{cccccc}
$(k_{\scriptscriptstyle\rm F})^{\scriptscriptstyle 0}$ 
& $\rho^{\scriptscriptstyle 0}_{\scriptscriptstyle N}$ 
& $M_{\scriptscriptstyle N,0}^*/M^*_{\scriptscriptstyle N}$ &
  $e_{\scriptscriptstyle 0}$ & $K_{\scriptscriptstyle 0}$ \\
\hline
1.3\,fm$^{-1}$ & 0.1484\,fm$^{-3}$ & 0.60 & $-16.1$\,MeV & 250\,MeV \\
\end{tabular}
\label{tab:one}
\end{table}
%
%

%
\newpage
%
\section*{Figure captions}
\global\firstfigfalse
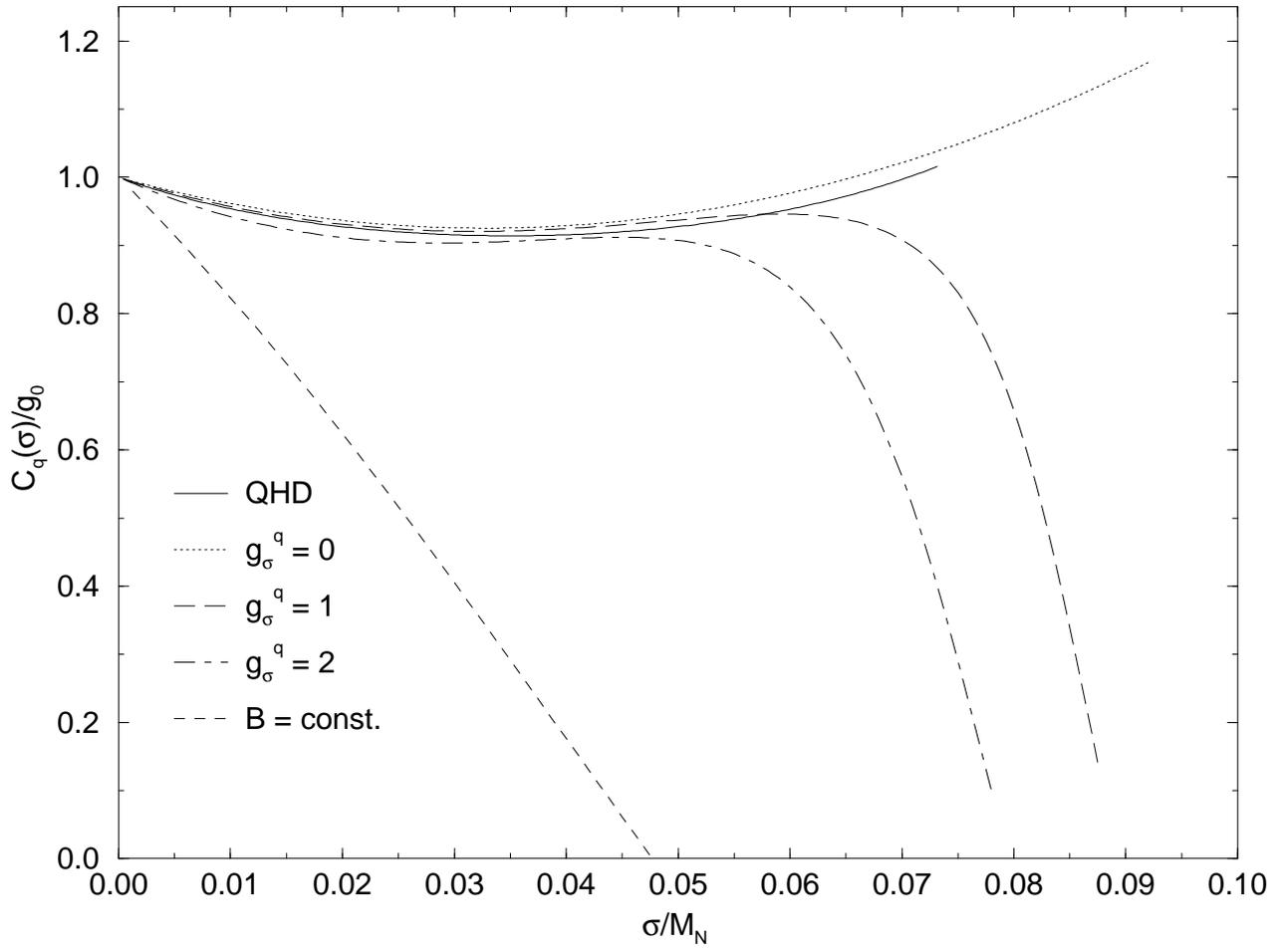
\begin{figure}[tbhp]
\caption{
Normalized effective coupling $C_q(\sigma)$ for the model MQMC$_{\rm A}$,
calculated with $\kappa=4$ and for various values of $g_\sigma^q$.
The corresponding quantity for QHD and for the original QMC model 
($B=const.$) is also indicated.
}
\label{fig:effcoup}
\end{figure}
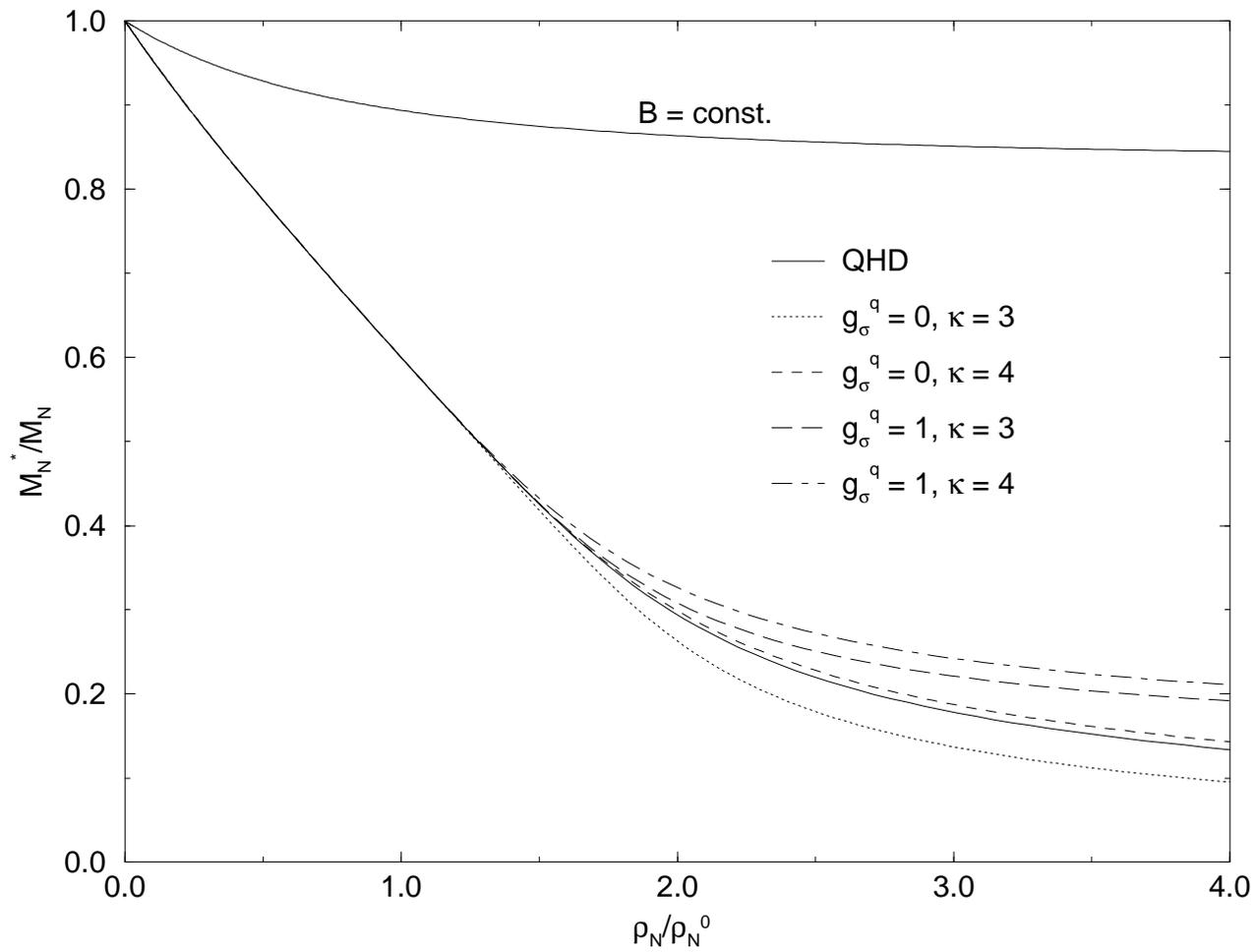
\begin{figure}[tbhp]
\caption{
Effective nucleon mass as a function of the density for the model
MQMC$_{\rm A}$. Results for various values of $\kappa$ and $g_\sigma^q$
are compared with QHD and the original QMC model ($B=const.$).
}
\label{fig:mstardc}
\end{figure}
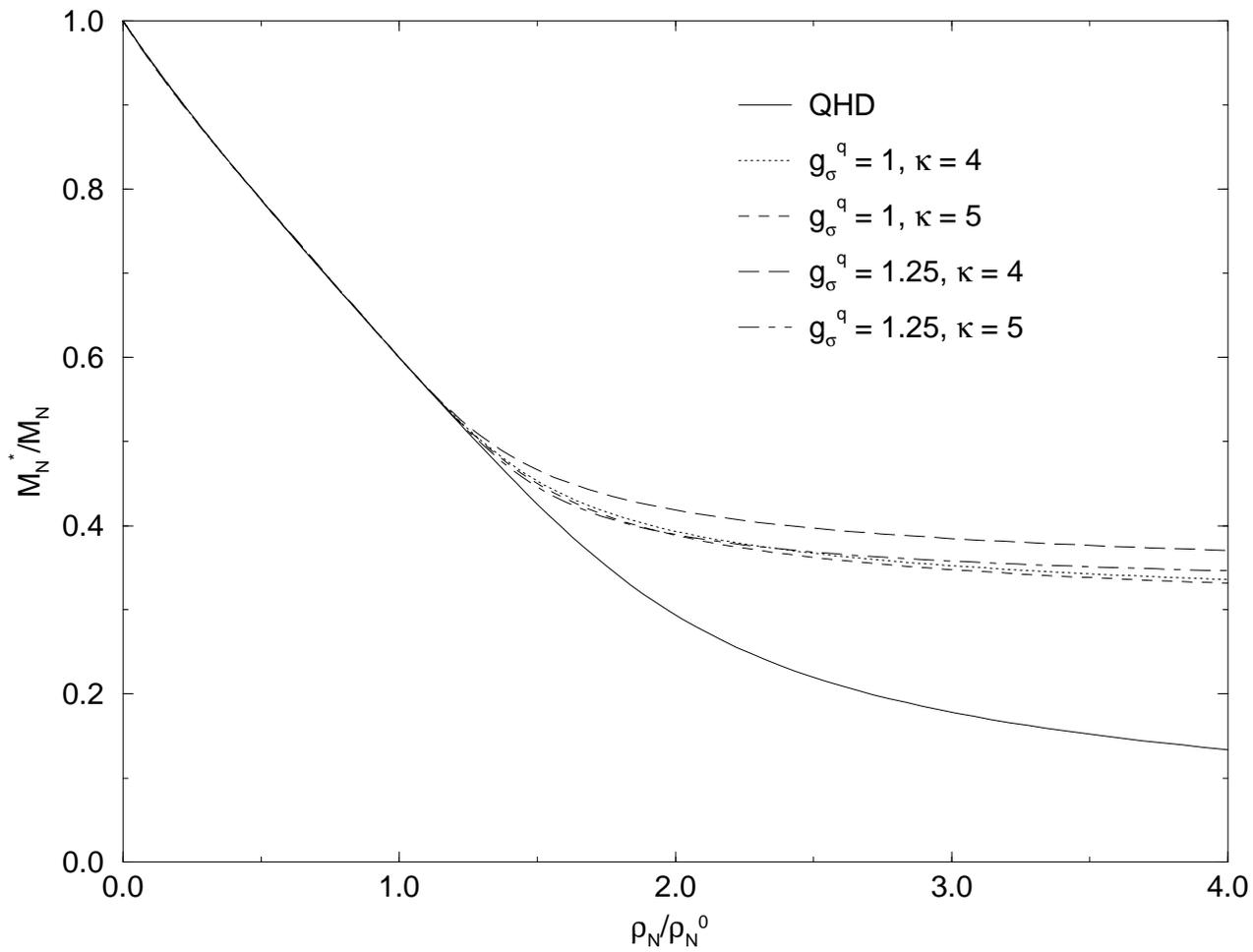
\begin{figure}[tbhp]
\caption{
Effective nucleon mass as a function of the density for the model 
MQMC$_{\rm B}$.
Results for different parameter sets are compared with QHD.
}
\label{fig:mstarsc}
\end{figure}
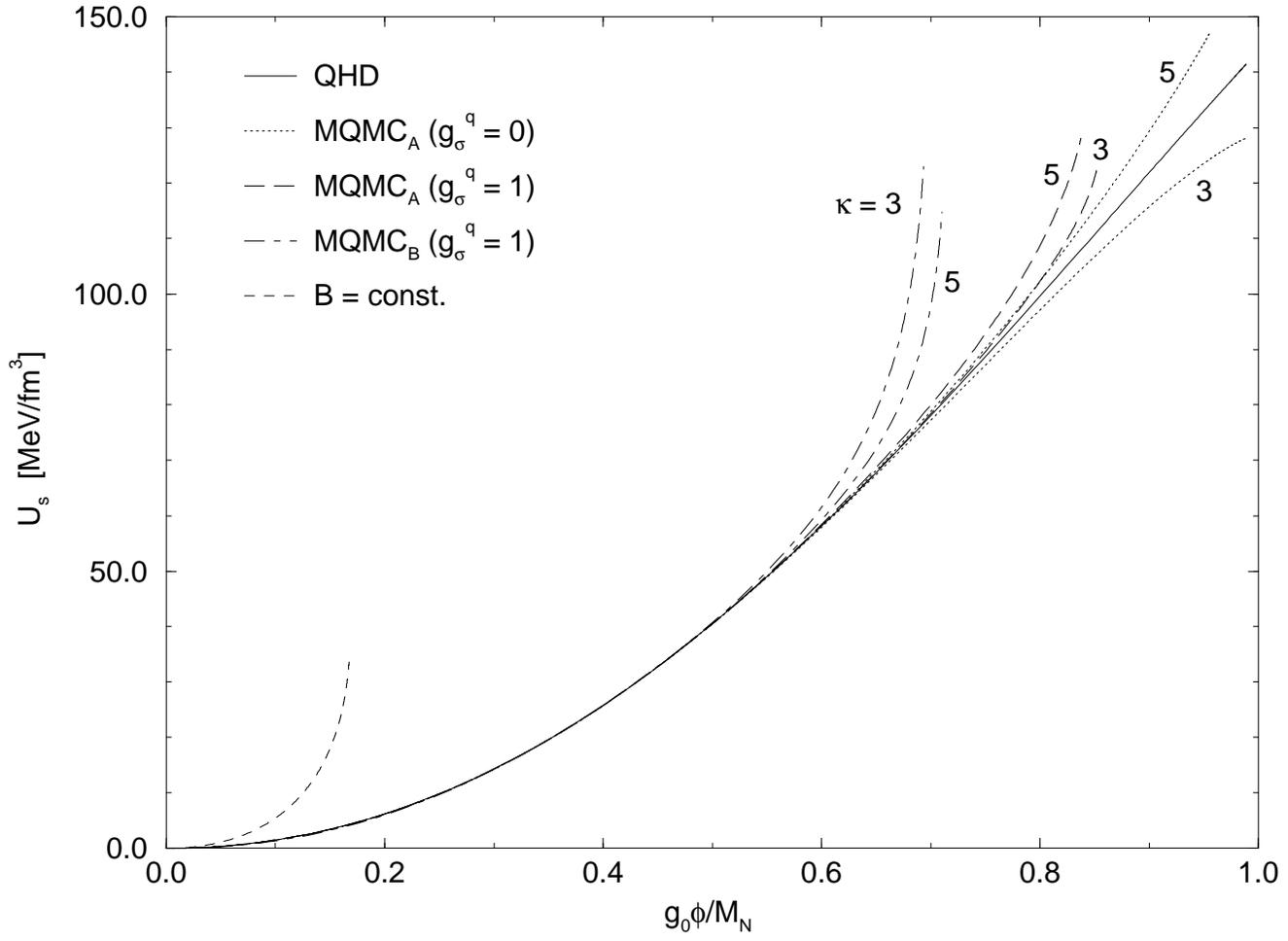
\begin{figure}[tbhp]
\caption{
Predicted nonlinear scalar potential as a function of the transformed
scalar field $g_0 \phi=M_N-M_N^*$.
The dotted and dashed curves correspond
to MQMC$_{\rm A}$ with $g_\sigma^q = 0$ and $g_\sigma^q = 1$ respectively.
The dotted-dashed curves indicate the result for MQMC$_{\rm B}$. 
In addition the result for the original QMC model ($B=const.$) and the
QHD potential is also indicated.
}
\label{fig:pot}
\end{figure}
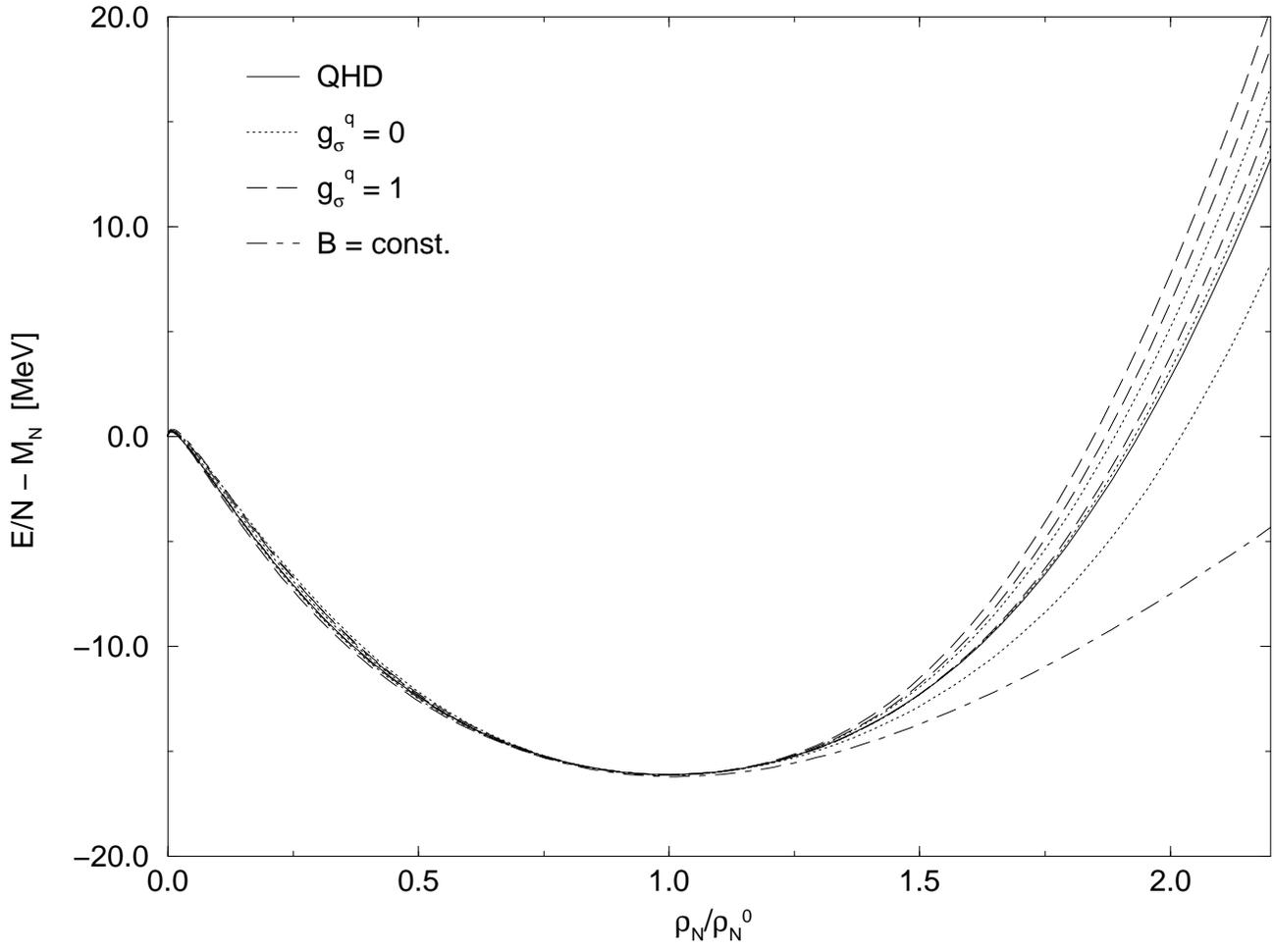
\begin{figure}[tbhp]
\caption{
Binding energy as a function of the density for MQMC$_{\rm A}$.
The dotted and dashed curves are calculated for $\kappa= 3,4,5$ from the bottom 
to the top. We also show the binding energy for the original QMC model 
($B=const.$) and for QHD.
}
\label{fig:eos}
\end{figure}
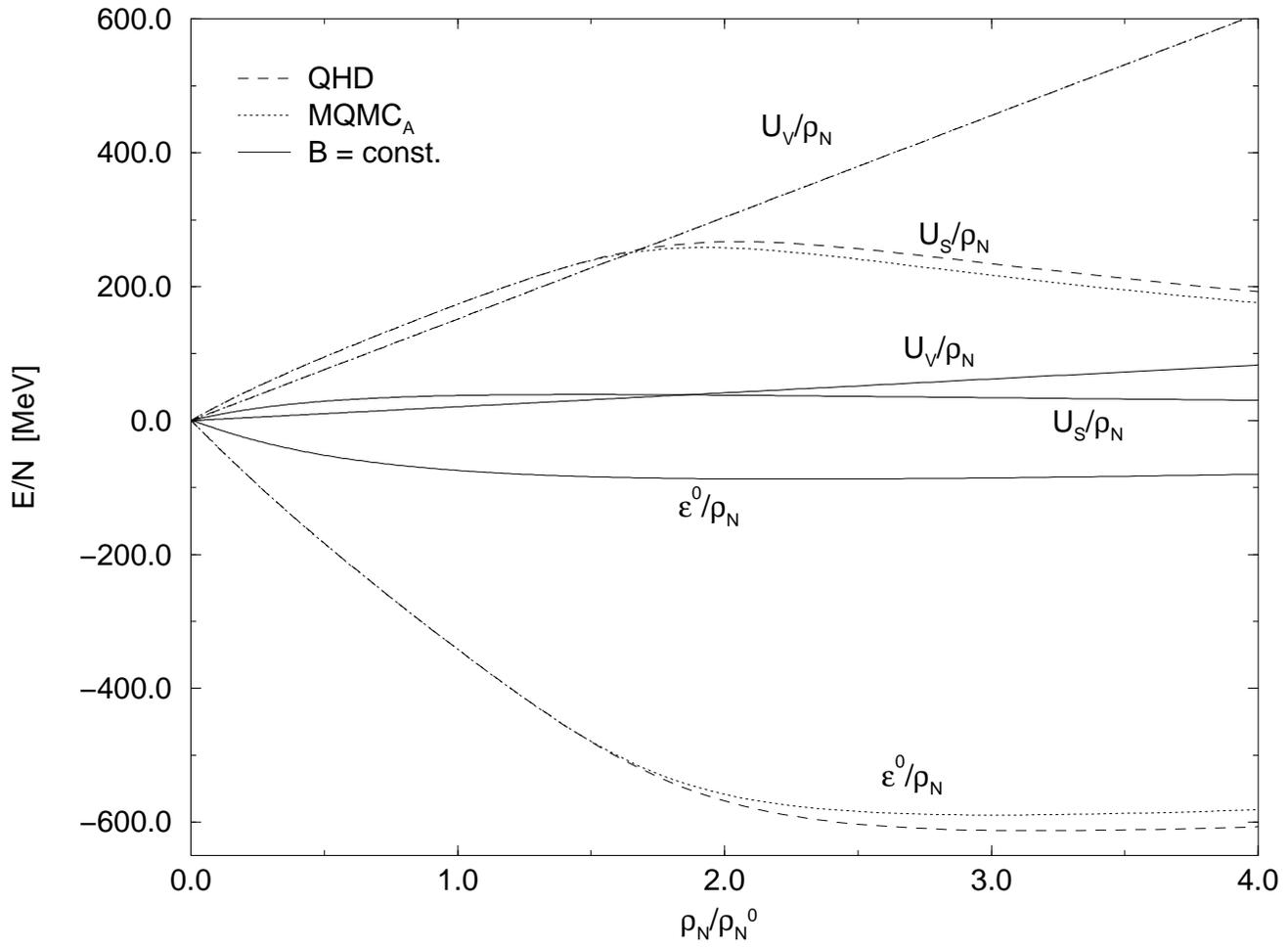
\begin{figure}[tbhp]
\caption{
Kinetic energy ${\cal E}^0/\rho_N$, 
vector potential $U_{\rm v}/\rho_N$
and scalar potential $U_{\rm s}/\rho_N$ as a function of the density.
Results for MQMC$_{\rm A}$, QHD and the original QMC model ($B=const.$)
are shown.
The parameters for MQMC$_{\rm A}$ are $\kappa= 3$ and $g_\sigma^q = 1$.
}
\label{fig:sat}
\end{figure}
\begin{figure}[tbhp]
\caption{
Bag constant as a function of the effective nucleon mass
calculated for $\kappa=4$ and for different quark-meson couplings $g_\sigma^q$.
In part (a) we show the results for MQMC$_{\rm A}$.
Here the curve $(M_N^*/M_N)^4$ corresponds to $g_\sigma^q=0$.
Part (B) indicates the results for MQMC$_{\rm B}$.
}
\label{fig:bag}
\end{figure}
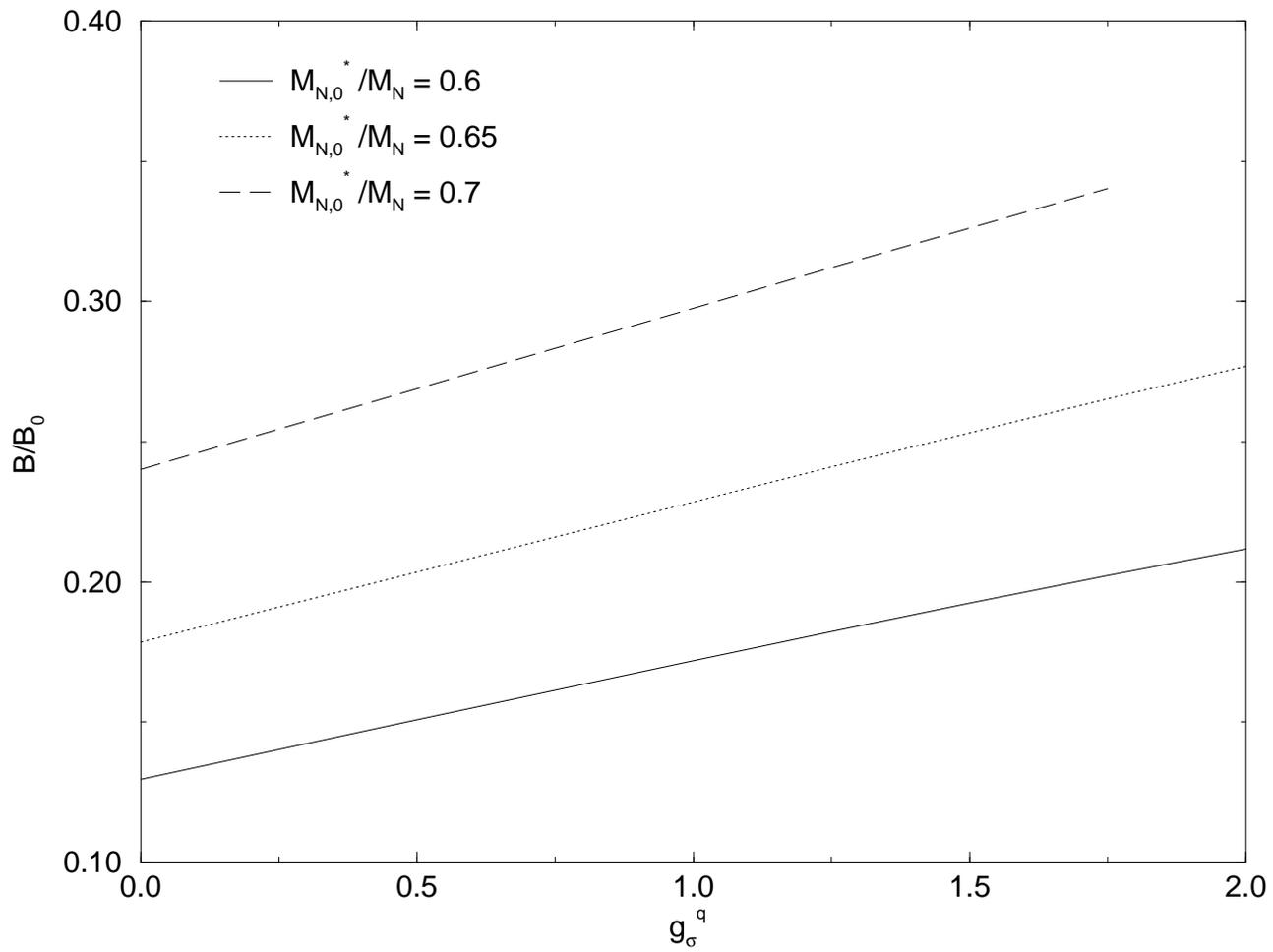
\begin{figure}[tbhp]
\caption{
Bag constant at the saturation point for MQMC$_{\rm A}$ as a 
function of the quark-meson coupling $g_\sigma^q$. 
The individual curves correspond to different values of the
nucleon mass at equilibrium.
The bag constant at equilibrium does not
depend on $\kappa$ and the compressibility.
}
\label{fig:bagsat}
\end{figure}
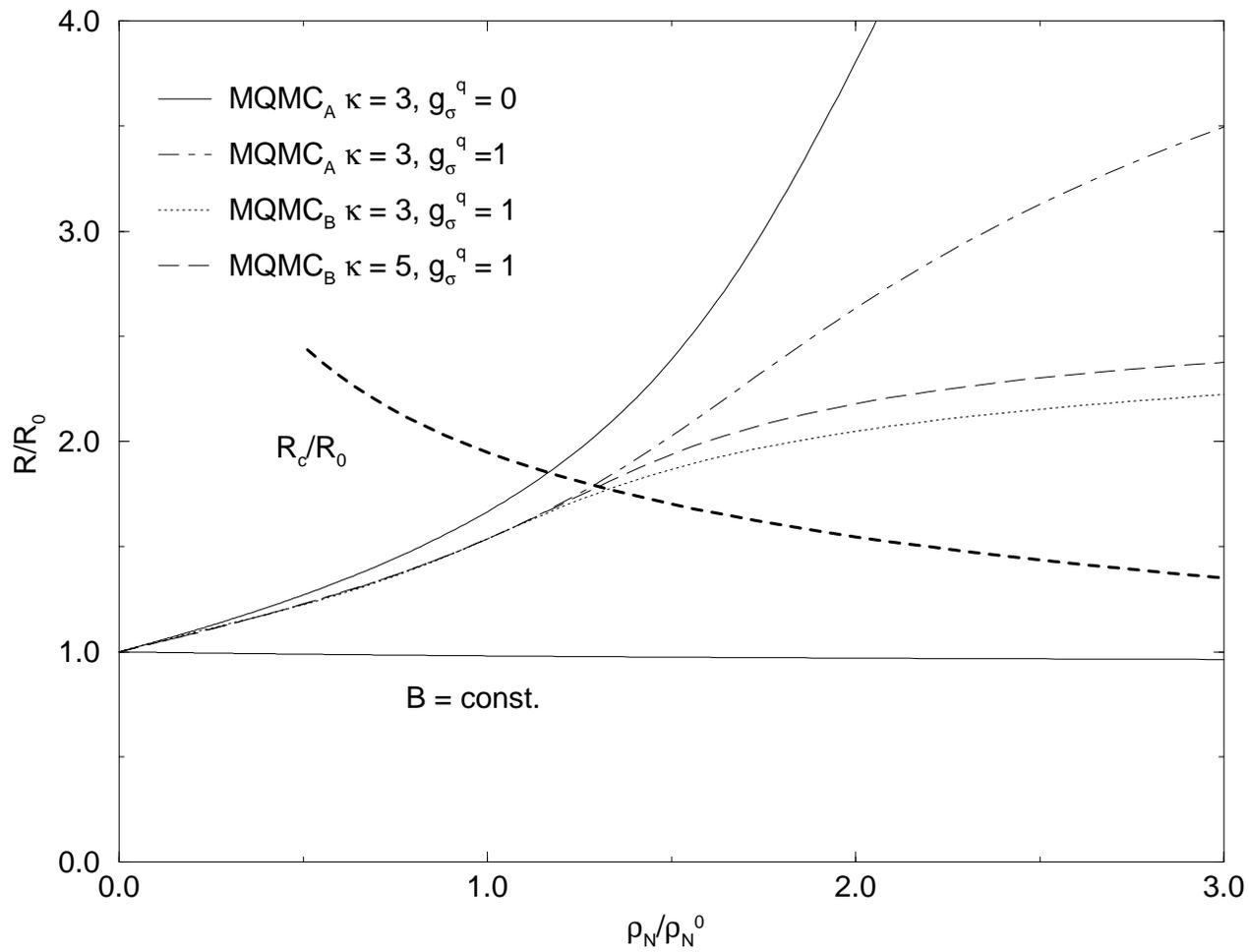
\begin{figure}[tbhp]
\caption{
Bag radius as a function of the density. At the critical radius 
$R_c =(3/4\pi\rho_N)^{1/3}$
the individual nucleon bags start overlapping.
}
\label{fig:radius}
\end{figure}
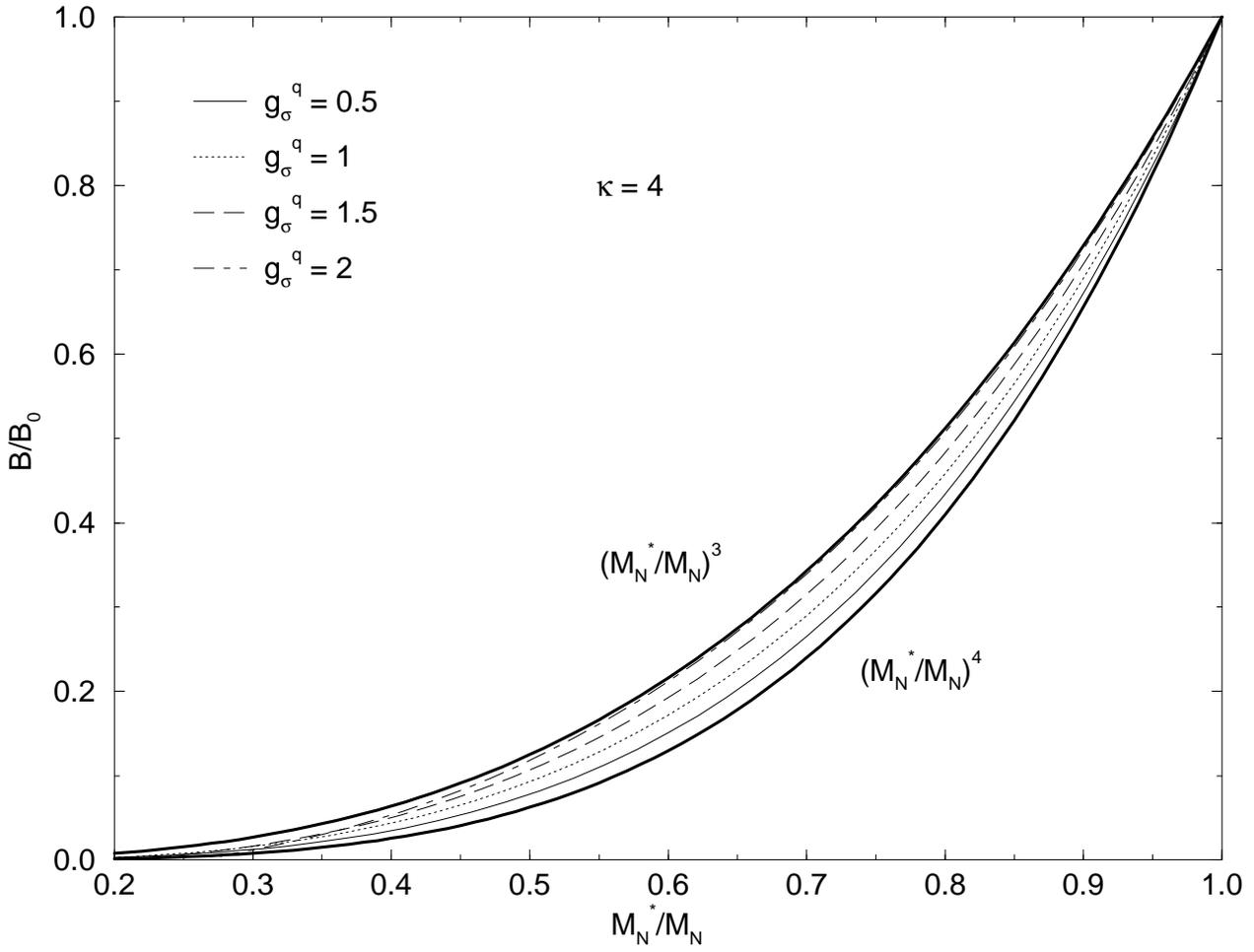
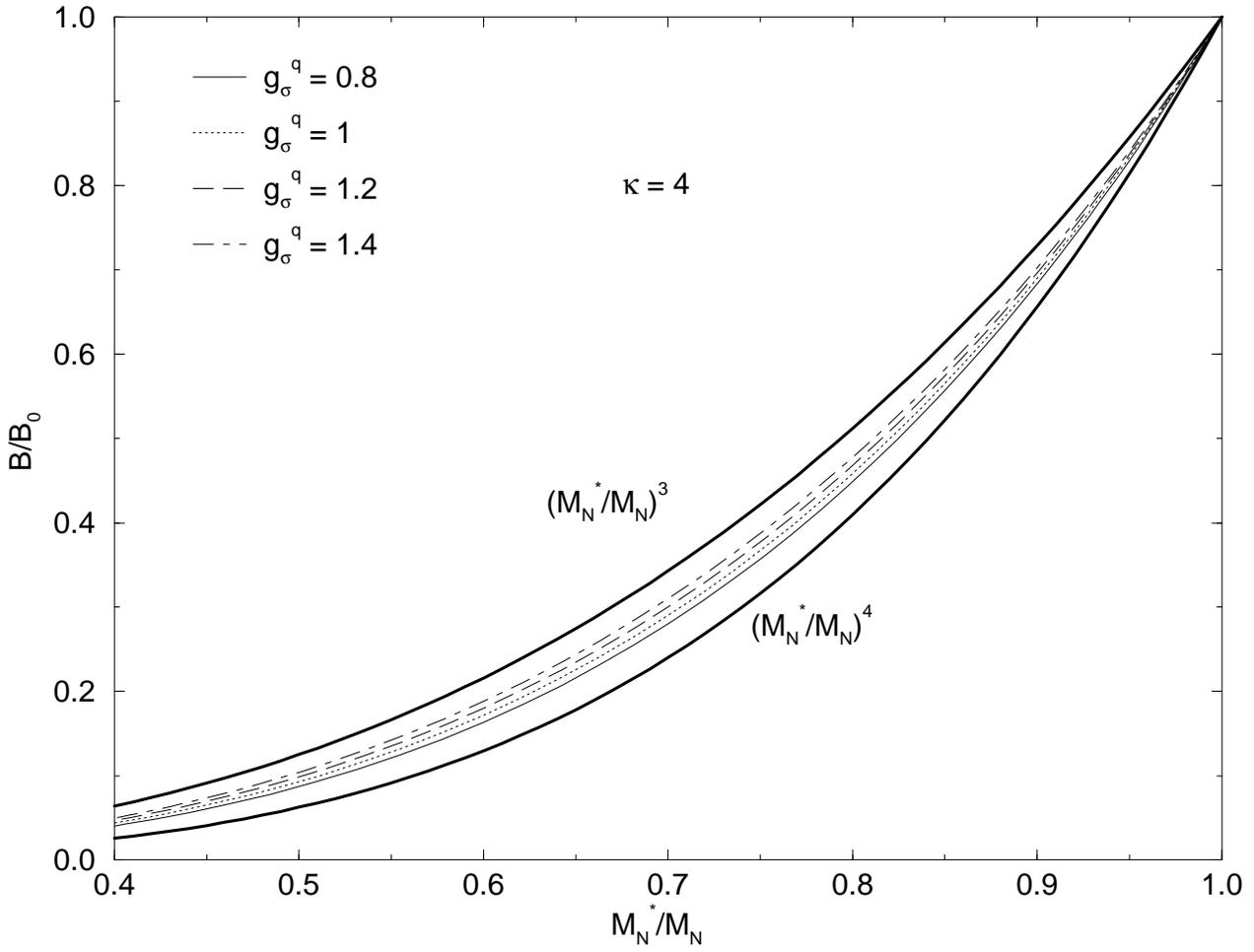
\end{document}